\documentclass{article}

\usepackage{amsmath}

\usepackage{graphicx}
\usepackage{dcolumn}
\usepackage{bm}
\usepackage{color}
\usepackage{epstopdf}
\usepackage[utf8]{inputenc}

\title{A strong polarizing field in thin-film paraelectrics}
\author{Chiara Gattinoni and Nicola A. Spaldin\\
Department of Materials, ETH Zurich, CH-8093 Zürich, Switzerland}
\date{}

\usepackage{graphicx}

\begin{document}

\maketitle

\begin{abstract}
 
The surface charge associated with the spontaneous polarization in ferroelectrics is well known to cause a depolarizing field that can be particularly detrimental in the thin-film geometry desirable for microelectronic devices \cite{Wurfel/Batra:1973,Dawber/Rabe/Scott:2005}. Incomplete screening of the surface charge, for example by metallic electrodes or surface adsorbates, can lead to the formation of domains \cite{Lichtensteiger_et_al:2007}, suppression or reorientation of the polarization \cite{Junquera/Ghosez:2003}, or even stabilization of a higher energy non-polar phase \cite{Mundy_et_al:2020}. A huge amount of research and development effort has been invested in understanding the depolarizing behavior and minimizing its unfavorable effects.
Here we demonstrate the opposite behavior: A strong {\it polarizing field}, which derives from the same physics as the depolarizing field in ferroelectrics, but which drives thin films of materials that are centrosymmetric and paraelectric in their bulk form into a non-centrosymmetric, polar state. We illustrate the behavior using density functional computations for perovskite-structure potassium tantalate, KTaO$_3$, which is of considerable interest for its high dielectric constant, proximity to a quantum critical point and superconductivity. We then provide a simple recipe to identify whether a particular material and film orientation will exhibit the effect, and develop an electrostatic model to estimate the critical thickness of the induced polarization in terms of well-known material parameters. Our results provide practical guidelines for exploiting the electrostatic properties of thin-film ionic insulators to engineer novel functionalities for nanoscale devices. 

\end{abstract}

\section{Introduction}

In addition to the bound surface charge arising from the bulk spontaneous polarization in ferroelectrics, surface charge can occur in thin films of {\it centrosymmetric} insulators that contain charged ionic layers perpendicular to the surface plane. This layer surface charge, $\sigma^{\text{surf}}_{\text{layer}}$, can be conveniently formalized in terms of a bulk layer polarization, as
\begin{equation}
\sigma^{\text{surf}}_{\text{layer}} = \vec{P}_{\text{layer}} \cdot \vec{n} \quad ,
\label{Eqn:surface_charge}
\end{equation}
where $\vec{P}_{\text{layer}}$ is the dipole moment per unit volume of the unit cell that tiles the semi-infinite plane containing the surface in the centrosymmetric structure \cite{Stengel:2011}. In ferroelectric materials with charged ionic layers, the layer surface charge can add to or subtract from the surface charge resulting from the spontaneous polarization, causing a complex interplay between polarization orientation and surface electrochemistry \cite{Yang_et_al:2017}. Recently, it was even demonstrated that, for the case of multiferroic bismuth ferrite, BiFeO$_3$, the surface charges from the layer and spontaneous polarizations almost exactly compensate for appropriate choices of surface termination and spontaneous polarization orientation, leading to electrostatically stable ferroelectric thin films with zero depolarizing field even in the absence of metallic electrodes \cite{Efe/Spaldin/Gattinoni:2020,Spaldin_et_al:2021}. In this work we reveal an even more dramatic consequence of this interplay between layer and spontaneous polarization, showing computationally, for the first time to our knowledge, that the surface charge arising from the layer polarization can {\it induce} ferroelectric-like polarization in a thin film of a material that is paraelectric in its bulk ground state.

\begin{figure}[ht]
    \centering
    \includegraphics[width=0.8\linewidth, trim={0 10cm 0 0},clip]{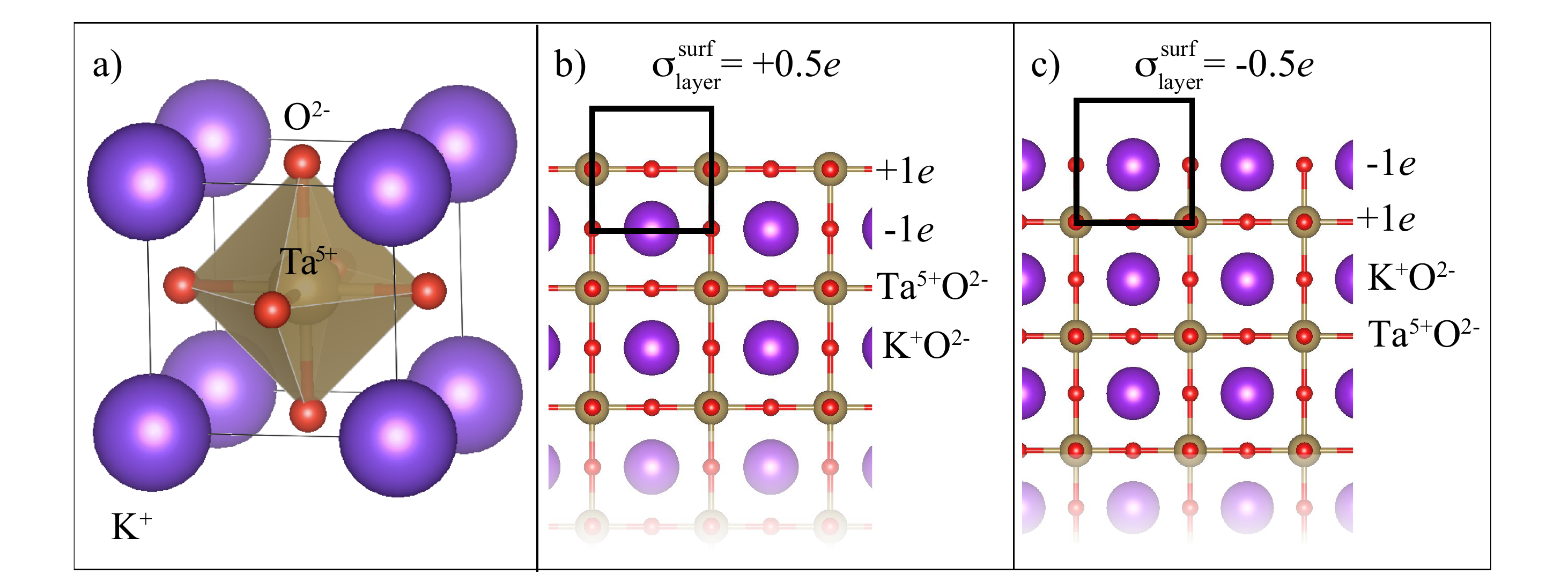}
    \caption{Structure of bulk and thin-film KTaO$_3$. a) The cubic perovskite unit cell of bulk KTaO$_3$. K, Ta and O, with formal charges +1, +5 and -2 electronic charges, $e$, are shown in purple gold and red respectively. b) and c) Structures of the upper portions of [001]-oriented KTaO$_3$ slabs with flat TaO$_2$ (b) and KO (c) surfaces. The formal charges of the TaO$_2$ (+1$e$) and KO (-1$e$) layers are shown next to the layers. The black rectangles indicate the unit cells that tile the semi-infinite plane and are used to calculate the surface charge, $\sigma^{\text{surf}}_{\text{layer}}$, using Eqn.~\ref{Eqn:surface_charge}.  We see that, for the TaO$_2$ surface, this unit cell has a charge of -1$e$ at position 0 and +1$e$ at position 0.5 of the lattice constant along the [001] direction, so its surface charge $\sigma^{\text{surf}}_{\text{layer}}$ and corresponding dipole per unit volume $P^{\text{[001]}}_{\text{layer}} = 0.5 e$ per unit-cell area, whereas for the KO surface, $\sigma^{\text{surf}}_{\text{layer}} = P^{\text{[001]}}_{\text{layer}} = -0.5 e$ per unit-cell area.  \label{fig:kto_structure}}
\end{figure}

We choose [001]-oriented perovskite-structure potassium tantalate, KTaO$_3$ (Fig.~\ref{fig:kto_structure} a), as our model system to demonstrate the behavior. First, its formal ionic charges, K$^+$, Ta$^{5+}$, O$^{2-}$, lead to charged (001) layers 
and correspondingly to a non-zero bulk layer polarization, with substantial positive and negative surface charges for (001) TaO$_2$  and KO surfaces (Figs.~\ref{fig:kto_structure} b and c) respectively. Second, there has been considerable interest in its surface properties, with the layer surface charge shown to cause complex surface reconstructions, including O adatoms or vacancies, stripes of alternating terminations, metallicity, and polar distortions~\cite{DeaconSmith_et_al:2014,setvin_science_2018,wang_pccp_2018,Zhao/Selloni:2019} depending on the external conditions. 
Reports of a surface two-dimensional electron gas~\cite{King_et_al:2012,SantanderSyro_et_al:2012} and superconductivity with critical temperature depending on the choice of surface plane are particularly intriguing~\cite{Ueno_et_al:2011,chen_prl_2021,liu_science_2021}. Finally, while it exhibits the centrosymmetric ideal cubic structure at all temperatures, its dielectric constant increases strongly as temperature is reduced, suggestive of quantum paraelectric behavior \cite{Barrett:1952,Wemple:1965, Rowley_et_al:2014}. As a result, the internal energy difference between paraelectric and ferroelectric states is small, making it particularly suitable for demonstrating the effect that we describe. 

\section{Methods}

Our calculations were performed using density functional theory (DFT) as implemented in the plane-wave code Quantum Espresso~\cite{Gianozzi_et_al:2009}. We used the PBE~\cite{pbe} parameterization of the generalized gradient approximation (GGA) to describe the exchange-correlation functional and replaced core electrons by GBRV pseudopotentials~\cite{gbrv_pseudopotentials}. 

In order to accurately model the small energy and structural differences between paraelectric and ferroelectric KTaO$_3$ we sampled the five-atom unit cell using a $24 \times 24 \times 24$
Monkhorst-Pack grid of $k$-points and a kinetic energy cut off of $75$ Ry for the wave-function and $750$ Ry for the charge density.
We set the force convergence threshold for structural optimization to $10^{-4}$ eV/\AA. Spin-orbit coupling was not included explicitly. 
The optimized lattice constant for the metastable cubic paraelectric unit cell was obtained by fixing the ions at the high symmetry positions and fitting the Birch-Murnaghan equation to the energy obtained by single point calculations at a range of lattice parameters.
For ferroelectric unit cell we instead minimized the Hellmann-Feynman stresses  and forces to obtain the lowest-energy structure.

We simulated slabs with $(1 \times 1)$ area and with thicknesses between four and 15 unit cells using periodic boundary conditions in the (001) plane and a separation of $15$ \AA\ of vacuum between periodic images along the [001] direction. All slabs had KTaO$_3$ stoichiometry, and so were not symmetric, consisting of one flat KO and one flat TaO$_2$ surface. 
For the slab calculations, we used a kinetic energy cutoff of $60$ Ry for the wave-function and $600$ Ry for the charge density, and set the force convergence threshold for structural optimizations to $10^{-4}$ eV/\AA.
An $8 \times 8 \times 1$ Monkhorst-Pack grid of $k$-points was used for all slabs.
We calculated the spontaneous polarization of the bulk material, and the layer-by-layer induced polarizations in the slabs by multiplying the displacement of each ion from its high symmetry position by its Born effective charge value taken from Ref.~\cite{Cabuk:2010} and normalizing to our calculated volume.

\newpage

\section{Results}

\subsection{Bulk KTaO$_3$}

Our calculated lattice parameter for cubic paraelectric KTaO$_3$, using the computational settings given in the Methods section, is $4.0276$ \AA, which is a slight overestimate of the experimental room-temperature lattice parameter of $3.989$ \AA\ \cite{Vousden:1951}. Full structural relaxation yields a slightly tetragonal unit cell, with 
$a= b = 4.0259$ \AA\, and $c= 4.031$ \AA, 
and a small spontaneous polarization of 11 $\mu$C/cm$^2$ along the [001] direction. The energy difference between the paraelectric and ferroelectric structures is tiny,  $0.2$ meV, consistent with the observed quantum paraelectric behavior. (We note that the energy difference between the paraelectric and ferroelectric structures is unusually sensitive to the choice of computational settings, consistent with the observed quantum paraelectric behavior). Our calculated band gap is 2.18 eV, underestimating the measured optical band gap of $\sim$3.6 eV. 

\begin{figure}[ht]
    \centering
    \includegraphics[width=0.6\linewidth, trim={0 0 5cm 0},clip]{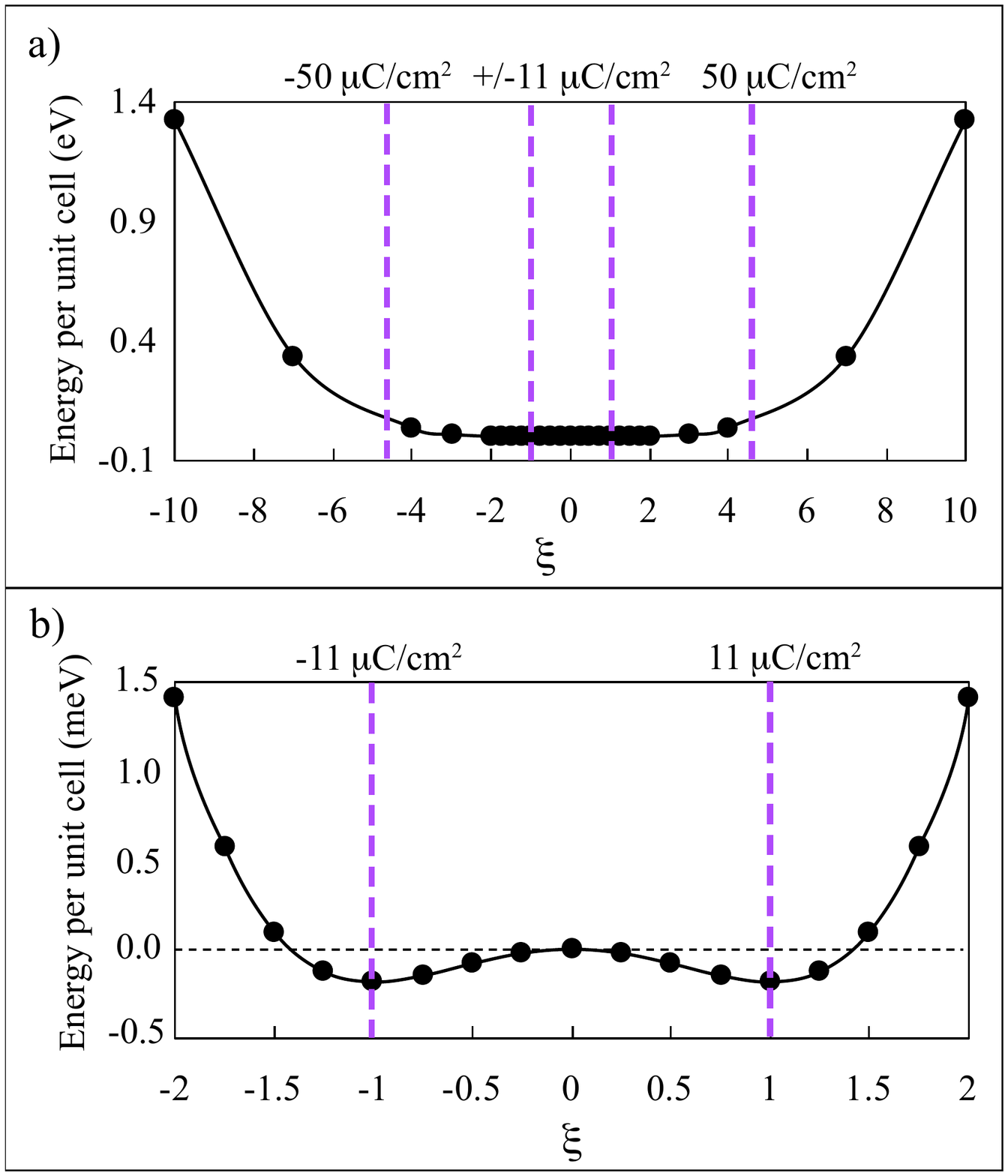}
    \caption{Calculated energy per unit cell as a function of the polar distortion amplitude $\xi$ for bulk KTaO$_3$. The zero of energy is set to that of the centrosymmetric structure with the lattice constants of the fully relaxed ferroelectric structure. a) shows the energy over a large range of distortion amplitudes, $-10 < \xi < 10$; b) is a close up of the range $-2 < \xi < 2$. Note the different scales on the energy axes in the two cases. \label{fig:kto_double_well}}
\end{figure}
Fig.~\ref{fig:kto_double_well} shows the calculated energy per formula unit as a function of the amplitude of the polar distortion, $\xi$, over large (a)  and small (b) ranges of $\xi$ (note the different units on the energy axis). Here $\xi=1$, with spontaneous polarization 11 $\mu$C/cm$^2$, corresponds to the fully relaxed structure and $\xi=0$ corresponds to the structure with the atoms at their centrosymmetric high-symmetry positions, which we set to have energy zero. The atomic positions are interpolated linearly between their values at $\xi=0$ and $\xi=1$, and extrapolated linearly to larger values of $\xi$, with all three lattice parameters fixed, for all values of $\xi$, to those of the fully relaxed ferroeletric unit cell. The $\xi$ values corresponding to polarizations of 11 $\mu$C/cm$^2$ (at the bottom of the double well) and 50 $\mu$C/cm$^2$ (to be discussed later) are indicated with the vertical dashed purple lines. (Note that changes in the details of the calculation, such as using the lattice parameter of the paralectric unit cell, or relaxing the lattice axis in the direction of the polar distortion lead to small changes in the shape of the double well. The exact choice of procedure does not, however, affect the findings of this paper).
 As expected, our zero kelvin DFT calculations, which do not include the quantum mechanical zero point energy, give a characteristic ferroelectric double well potential  with a very small energy barrier between the oppositely polarized states. Subsequent solution of the Schr\"odinger equation using this potential has its lowest energy eigenvalue above the height of the barrier \cite{Esswein/Spaldin:2021}. The DFT description, while of course not capturing explicitly the suppression of ferroelectricity by quantum fluctuations in the low-temperature state \cite{Akbarzadeh_et_al:2004}, is therefore consistent with quantum paraelectric behavior. 

\newpage

\subsection{KTaO$_3$ thin films}

\begin{figure}[ht]
    \centering
    \includegraphics[width=0.6\linewidth, trim={0 5cm 0 0},clip]{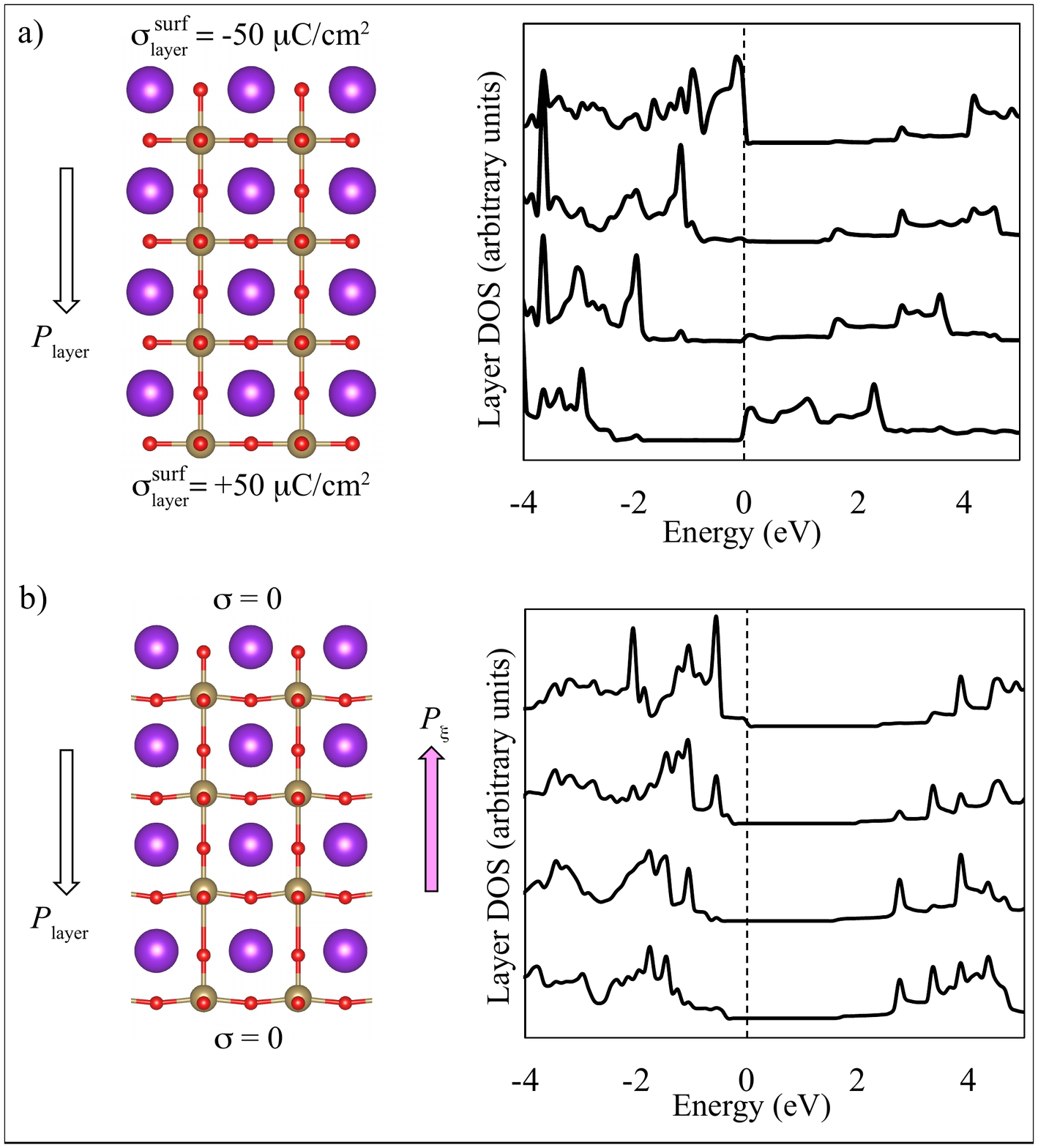}
    \caption{Structure (left) and layer-by-layer density of states (right) for a) cubic paraelectric and b) fully relaxed ferroelectric-like four unit cell (001) KTaO$_3$ slab. The orientation of the layer polarization $P_{\mathrm{layer}}$, which gives rise to the layer surface charges $\sigma_{\text{layer}} = \pm 50 \mu$C/cm$^2$, is indicated with the white arrows. In a) the surface charge can only be compensated by the generation of electron-hole pairs across the band gap, which manifests as the band bending shown in the layer-by-layer densities of states (DOS) in the right panel. In b) the surface charge is compensated by the ferroelectric-like polarization $P_{\xi}$, which can be seen in the relative displacements of the ions, and which is indicated by the purple arrow. As a result of this compensating polarization, there is no internal electric field and correspondingly no band bending in the layer-by-layer density of states. \label{fig:kto_001}}
\end{figure}

Having established that our DFT setup provides a good description of bulk KTaO$_3$, we now turn to thin films. We begin by constructing a four-unit-cell thick stoichiometric slab of cubic paraelectric KTaO$_3$ as described in the Methods section, placing the negatively charged KO layer on the top and the positively charged TaO$_2$ layer on the bottom as shown in the left panel of  Fig.~\ref{fig:kto_001} a). In the paraelectric structure, the only contribution to the surface charge is that from the layer polarization, which, as derived in Figs.~\ref{fig:kto_structure} b) and c), has a value of -0.5 electrons per unit cell area on the KO surface and +0.5 electrons per unit cell area on the TaO$_2$ surface. This converts to 
$\pm$ 50 $\mu$C/cm$^2$, at our calculated bulk lattice constant of 4.026 \AA .  An additional polarization from a ferroelectric-like polar lattice distortion along the [001] direction, $P_{\xi}^{[001]}$, of $\mp$ 50 $\mu$C/cm$^2$ would therefore fully compensate the surface charge from the layer polarization.

Next, we fully relax the atomic positions and out-of-plane lattice constant, with the in-plane lattice constant fixed to our calculated bulk value (4.026 \AA). 
We find that the lowest energy structure, shown in the left panel of  Fig.~\ref{fig:kto_001} b), develops a large ferroelectric-like polarization, with an average dipole moment per unit volume, $P_{\xi}$, close in value to the layer polarization of 50 $\mu$C/cm$^2$. $P_{\xi}$ points in the opposite direction to $P_{\text{layer}}$, so that its associated bound surface charges compensate those from the layer polarization. We note that, while part of  $P_{\xi}$ originates from the 11 $\mu$C/cm$^2$ spontaneous polarization seen in our potential energy calculation of Fig.~\ref{fig:kto_double_well}, the majority is induced by the polarizing field caused by the uncompensated surface charges associated with the large layer polarization. 

The electrostatic force driving the development of the ferroelectric-like induced polarization can be seen in the calculated layer-by-layer densities of states in the right panels of Figure~\ref{fig:kto_001}. Fig.~\ref{fig:kto_001} a) shows the layer-by-layer density of states (DOS) across the slab in which each layer is constrained to adopt the high-symmetry paraelectric structure. The uncompensated layer surface charges result in a large internal electric field, which can be seen in the bending of the energy bands across the slab. Indeed, the field is so strong that the Fermi energy (at 0 eV) lies in the conduction band at the lower (positively charged) surface layer, and in the valence band at the upper (negatively charged) surface layer. This provides electrons in the conduction band of the bottom layer, which compensate the positive TaO$_2$ layer charge, and holes in the valence band of the top KO layer, to compensate the negative surface layer charge. Fig.~\ref{fig:kto_001} b) shows the analogous layer-by-layer density of states for the relaxed structure, in which $P_{\xi}$ compensates the layer polarization. In striking contrast to the paraelectric slab, the internal electric field and corresponding band bending are minimal.

\begin{figure}[ht]
    \centering
    \includegraphics[width=0.8\linewidth]{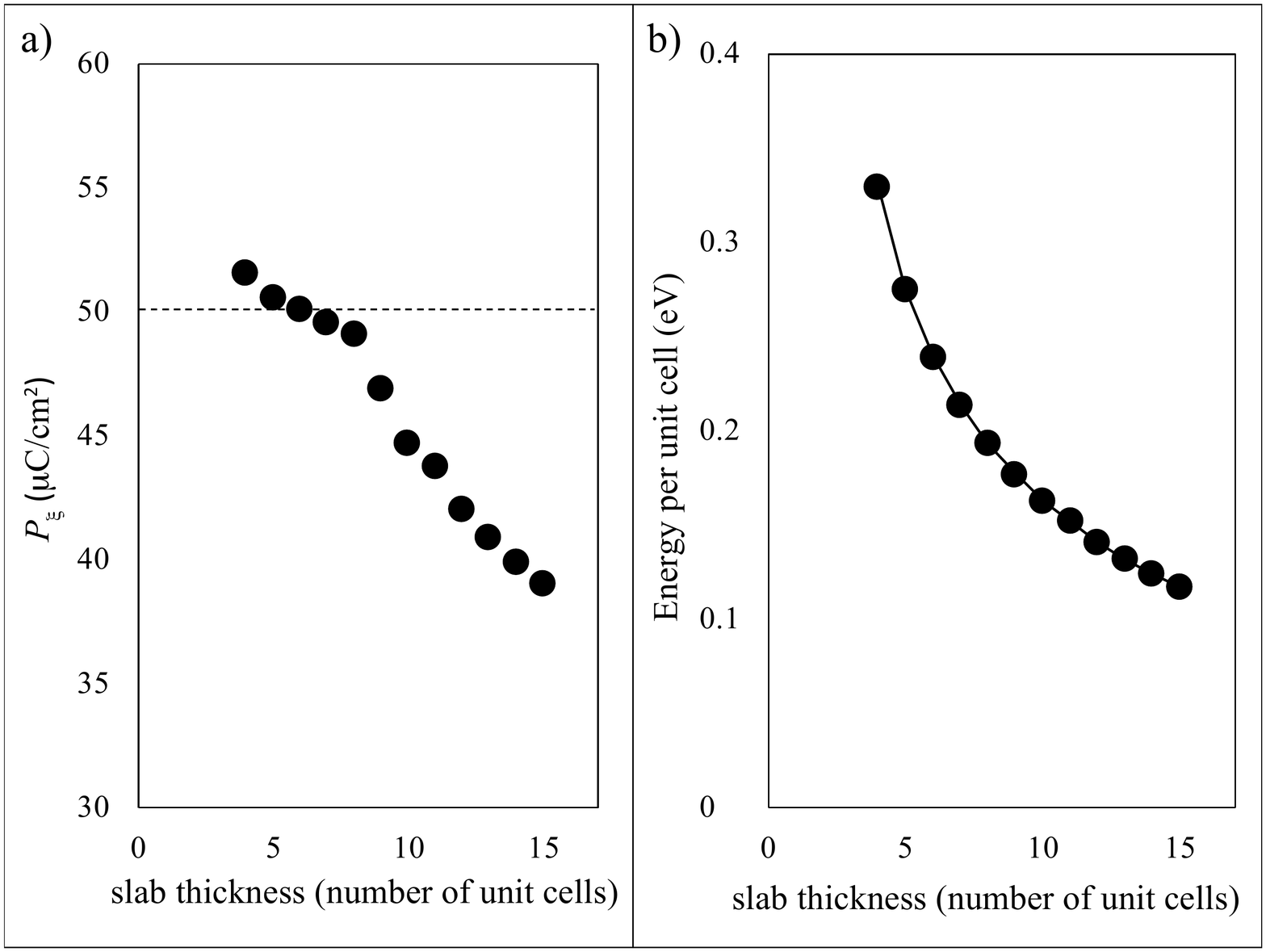}
    \caption{a) Ferroelectric-like polarization, $P_{\xi}$, and b) total energy per unit cell, for KTaO$_3$ slabs as a function of slab thickness, calculated using density functional theory. In panel b) the energy for an infinitely thick slab --- approximated as the energy of a bulk unit cell --- has been set as the zero. \label{fig:kto_thickness}}
\end{figure}

Finally, we repeat our DFT calculations for slabs of increasing thickness, and show our calculated ferroelectric-like polarization $P_{\xi}$ and energy per unit cell as a function of the slab thickness in Fig.~\ref{fig:kto_thickness}. $P_{\xi}$ (Fig.~\ref{fig:kto_thickness} a) remains approximately equal to $-P_{\text{layer}}$ until a thickness of eight unit cells, when it starts to decrease; the cross-over corresponds to the thickness at which compensating surface charges are generated via electron-hole excitation across the band gap (see Appendix~\ref{Appendix:pdos}).  Even at the largest thickness that is practical in our DFT calculations (15 unit cells), we find that the value of $P_{\xi}$ has dropped by only $\sim$ 10 $\mu$C/cm$^2$ from its maximum value. As expected, the energy per unit cell (Fig.~\ref{fig:kto_thickness} b), which is dominated by the surface energy from the broken chemical bonds at the two surfaces, decreases smoothly with thickness as the system tends to the bulk behavior.

\newpage
\subsection{Electrostatic Model}

In this last section, we present a simple model that allows for estimation of the polarization induced by a polarizing field in slabs of arbitrary thickness, given a particular lattice polarization, energy versus polar distortion profile, and screening mechanism. 
We include the following contributions to the energy in the model:
\begin{description}
\item[The bulk polarization energy.] We describe the internal energy per unit volume, $E_{\text{int}}$, as a function of the polarization arising from the ferroelectric-like distortion, $P_{\xi}$, in the usual Landau form as
\begin{equation} 
E_{\text{int}} = a P_{\xi}^2 + b P_{\xi}^4 \quad .
\end{equation}
The energy per unit surface area from this contribution is then $E_{\text{int}} d$, where $d$ is the film thickness. Fitting to our calculated energy versus $P_{\xi}$ profile over the range $-7 \leq \xi \leq 7$ gives values for the coefficients in this case of $a= - 2.816\times 10^{-3} $ meV cm$^4\mu$C$^{-2}$ and $b= 1.123 \times 10^{-5}$ meV cm$^8\mu$C$^{-4}$.

\item[The electrostatic energy from any unscreened polarization perpendicular to the slab surface.] The net polarization contributing to the electrostatic energy is the layer polarization, $P_{\text{layer}}$, minus any ferroelectric-like polarization, $P_{_{\xi}}$, that opposes it, further reduced by any charge accumulation on the surface from additional screening mechanisms. The net polarization causes a polarizing field,
\begin{equation}
    \mathcal{E}_{\text{pol}} = \frac{1}{\epsilon_0 \epsilon_r} (P_{\text{layer}} - P_{_{\xi}} -\sigma_{\text{scr}})
\end{equation}
which results in an electrostatic energy per unit area, $E_{\text{es}}$, of 
\begin{equation}
E_{\text{es}} = \frac{1}{2\epsilon_0 \epsilon_r} (P_{\text{layer}} - P_{_{\xi}} -\sigma_{\text{scr}})^2 d \quad ,
\end{equation}
where $\sigma_{\text{scr}}$ is the magnitude of the screening charge per unit area on each surface.
\item[The screening energy.] Following Ref.~\cite{Mundy_et_al:2020}, we assume that the screening mechanism is generation of electron-hole pairs across the gap, so that the screening energy per unit area, $E_{\text{scr}}$, is given by 
\begin{equation}
E_{\text{scr}} = \frac{\sigma_{\text{scr}}}{e}E_g \quad .
\end{equation}
Here $e$ is the size of the electronic charge and $E_g$ is the band gap (in eV). This screening mechanism is the only one available to a pristine system in vacuum, and provides an upper bound to the screening energy for a real system, in which lower energy screening processes associated with point defects or adsorbates \cite{Yang_et_al:2017} are likely available.
\end{description}
The total energy per unit area of the slab is the sum of these three contributions:
\begin{eqnarray}
E_{\text{tot}} & = & E_{\text{int}}d + E_{\text{es}} + E_{\text{scr}}  \nonumber \\
 & = & (a P_{_{\xi}}^2 + b P_{_{\xi}}^4) d + \frac{1}{2\epsilon_0 \epsilon_r} (P_{\text{layer}} - P_{_{\xi}} -\sigma_{\text{scr}})^2 d + \frac{\sigma_{\text{scr}}}{e}E_g \quad .
 \label{Eqn:Etot}
\end{eqnarray}
Note that the internal and electrostatic energies have the same dependence on slab thickness, $d$, and so in the absence of electron-hole excitations across the gap, $P_{\xi}$ is not thickness-dependent. 

\begin{figure}[ht]
    \centering
    \includegraphics[width=0.5\linewidth,trim={0  5cm 0 5cm}]{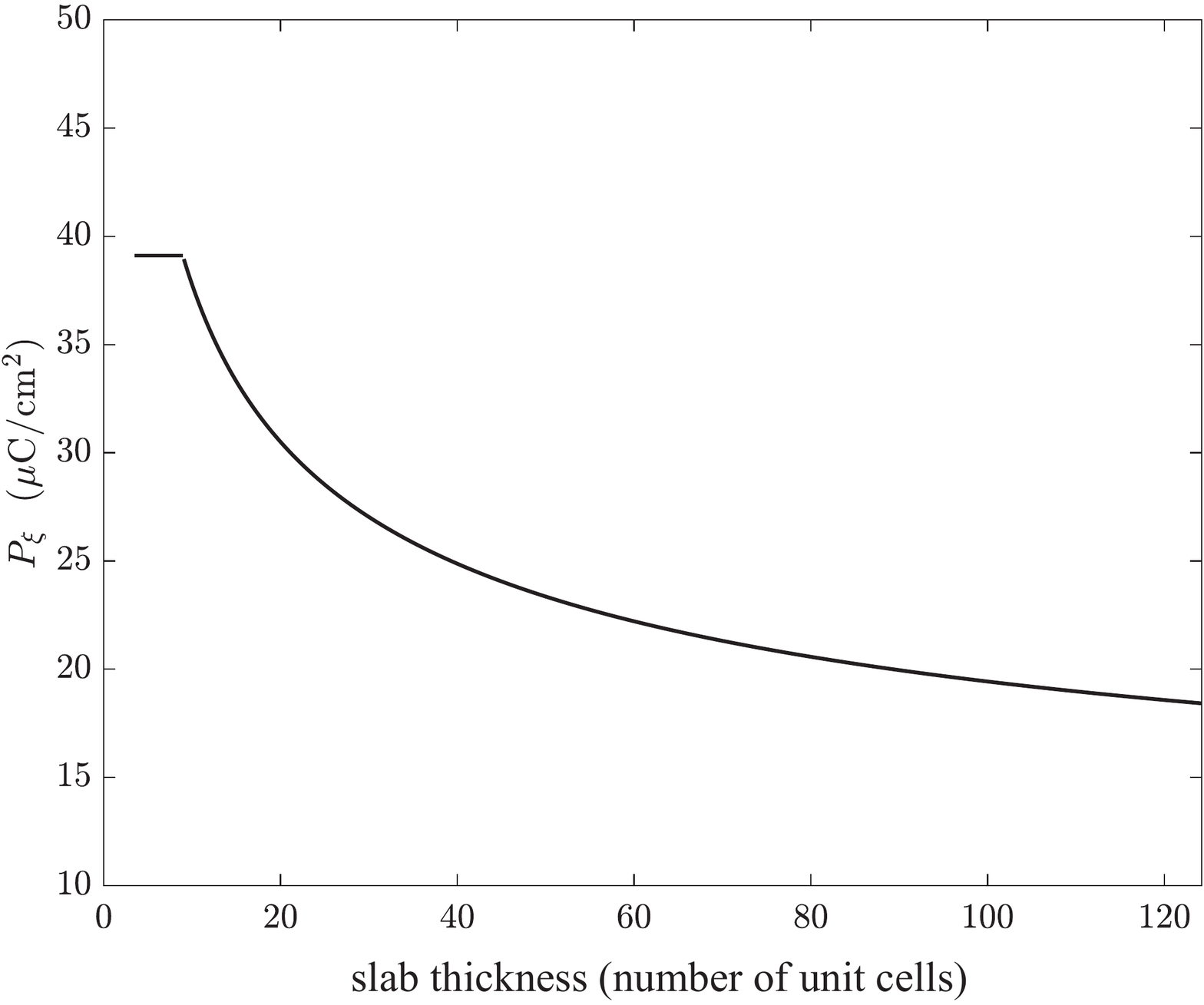}
    \caption{Ferroelectric-like polarization, $P_{_{\xi}}$, as a function of slab thickness, calculated using the simple electrostatic model of Eqn.~\ref{Eqn:Etot}.  \label{minPFE_vs_d_figure}}
\end{figure}

We then calculate the values of $P_{\xi}$ and $\sigma_{\text{scr}}$ that minimize this total energy for each value of $d$, noting that there is a critical thickness for the formation of screening electron-hole pairs because of the finite size of the band gap. The detailed derivation is provided in Appendix~\ref{Appendix:Etot} and the {\sc Matlab} code in Appendix~\ref{Appendix:Matlab}. Our calculated $P_{\xi}$ as a function of slab thickness is shown in Fig.~\ref{minPFE_vs_d_figure}, for the formal charge value of $P_{\text{layer}} = 50$ $\mu$C/cm$^2$, our calculated $E_g = 2.18$ eV, and a representative $\epsilon_r = 20$. As expected, for small slab thicknesses at which screening charges are not generated, $P_{\xi}$ remains constant at a value close to the layer polarization. Above a thickness of 8 unit cells, it becomes energetically favorable to form screening charges via excitation across the band gap, and $P_{\xi}$ begins to decrease. $P_{\xi}$ reaches its lower limit (the spontaneous polarization at the energy minima of Fig.~\ref{fig:kto_double_well}) only at thicknesses of several microns, indicating that the physics described here are valid for films far beyond the ultra-thin limit accessible using DFT. Increasing the value of the relative permittivity in the model decreases the ferroelectric-like polarization at small thicknesses and increases the thickness range before the electron-hole screening mechanism is activated; for $\epsilon_r = 100$ these values are $\sim$ 30 $\mu$C/cm$^2$ and 25 unit cells thickness respectively.

\section{Discussion}

In summary, we have shown that thin films of insulating materials with charged ionic layers can have an internal {\it polarizing field} that is strong enough to drive the formation of a ferroelectric-like induced polarization in otherwise paraelectric materials. This polarizing field derives from the same physics -- the drive to compensate bound surface charges -- as the \textit{depolarizing} field of ferroelectric materials, but manifests in the opposite way, by inducing polarization in the paraelectric phase rather than suppressing the polarization of a ferroelectric lattice.
The effect occurs when the paraelectric state has a surface charge associated with a component of the bulk layer polarization lattice~\cite{Stengel:2011} perpendicular to the surface normal, and when the internal energy cost to deform the ground-state paraelectric structure into a polar state is not large. We illustrated the behavior using [001]-oriented films of stoichiometric KTaO$_3$ with flat (001) surfaces, for which our DFT calculations yield a polar ground state up to the thickness limit (15 unit cells) accessible in our computations. The induced polarization is oriented pointing away from the TaO$_2^+$ and towards the KO$^-$ surface so that its associated surface charge partially compensates that from the charged ionic layers. Finally, we introduced a simple electrostatic model, which suggests that such a ferroelectric-like polarization is induced in KTaO$_3$ up to thicknesses of several microns, and which is easily generalizable to other materials. 

The case of KTaO$_3$ is particularly relevant in light of recent experiments suggesting an influence of the surface orientation on superconductivity. Using the method presented above to calculate the layer polarization and the corresponding surface charge, we see that for the high Miller-index planes, the surface charges associated with the bulk layer polarization for {\it flat} surfaces or interfaces follow the trend  $\sigma_{\text{surf}}^{\text{(111)}} > \sigma_{\text{surf}}^{\text{(110)}} > \sigma_{\text{surf}}^{\text{(001)}}$. Consistent with the need to screen the associated polar discontinuities, the surfaces readily form two-dimensional electron gases that have been well characterized using angle-resolved photoemission~\cite{Bruno_et_al:2019, King_et_al:2012, Bareille_et_al:2014}. (Note that an interface of KTaO$_3$ is electrostatically similar to a surface, provided that its neighbor is not a coherent I-V perovskite oxide, and so we use the term surface also for interfaces with systems that are not the vacuum). These surface two-dimensional electron gases become superconducting at low temperature, with the reported superconducting critical temperatures, $T_c$, different for the three orientations. Intriguingly, the (111) surfaces have the highest reported $T_c$ (2.2 K)~\cite{liu_science_2021} followed by 0.9 K for (110)~\cite{chen_prl_2021} and 50 mK for (001)~\cite{Ueno_et_al:2011} surfaces, following the same trend as the surface charges associated with the bulk layer polarization.
Detailed investigation of the structure of the surfaces of superconducting KTaO$_3$ samples in the three orientations is therefore of utmost interest. We note, in particular, that both the (110) and (111) surfaces are able to lower their surface charges by forming steps (in the (110) case an uncharged surface is even possible from such rearrangements), and that such behavior is likely.

While KTaO$_3$ is particularly suitable for illustrating the consequences of the polarizing field for the reasons discussed above, some degree of induced polarization is likely in thin enough films of all materials that have a surface charge associated with a layer polarization, with the thickness for which an induced polarization is maintained determined by the balance of contributions to the total energy given in Eqn.~\ref{Eqn:Etot}. Among perovskite-structure oxides, the flat (001) and (111) surfaces of III-III (with both A and B cations trivalent) and I-V (with monovalent A-site cations and pentavalent B-site cations) structures are all candidates. 

Finally, we note that the behavior we describe is robust to, and can even be slightly favored by, biaxial strain introduced through coherent heteroepitaxy, which modifies the balance of the contributions to the total energy by changing the unit-cell surface area and tuning the $a$ and $b$ parameters in $E_\text{int}$. In this context, the concepts discussed here could be particularly relevant for thin-film growth processes, with the fabrication of ultra-thin films in orientations corresponding to charged surfaces likely less prohibitive than previously expected. We hope that our work inspires experimental efforts to demonstrate, characterize and exploit the polarizing field in thin-film KTaO$_3$ and related materials.

\section{Acknowledgments}
This work was supported by the European Research Council (ERC) Grant Agreement No. 810451 (NAS) and the Marie Sklodowska–Curie Grant Agreement No. 744027 (CG), both part of the European Union’s Horizon 2020 research and innovation programmes, and by the ETH Zurich. Calculations were performed on the Euler cluster managed by the HPC team at ETH Zurich. NAS thanks Roy Smith for help with the {\sc Matlab} code.

\section{Appendices}

\appendix

\section{Layer-by-layer densities of states for eight- and nine-unit-cell thick slabs}
\label{Appendix:pdos}

\begin{figure}[ht]
    \centering
    \includegraphics[width=0.8\linewidth]{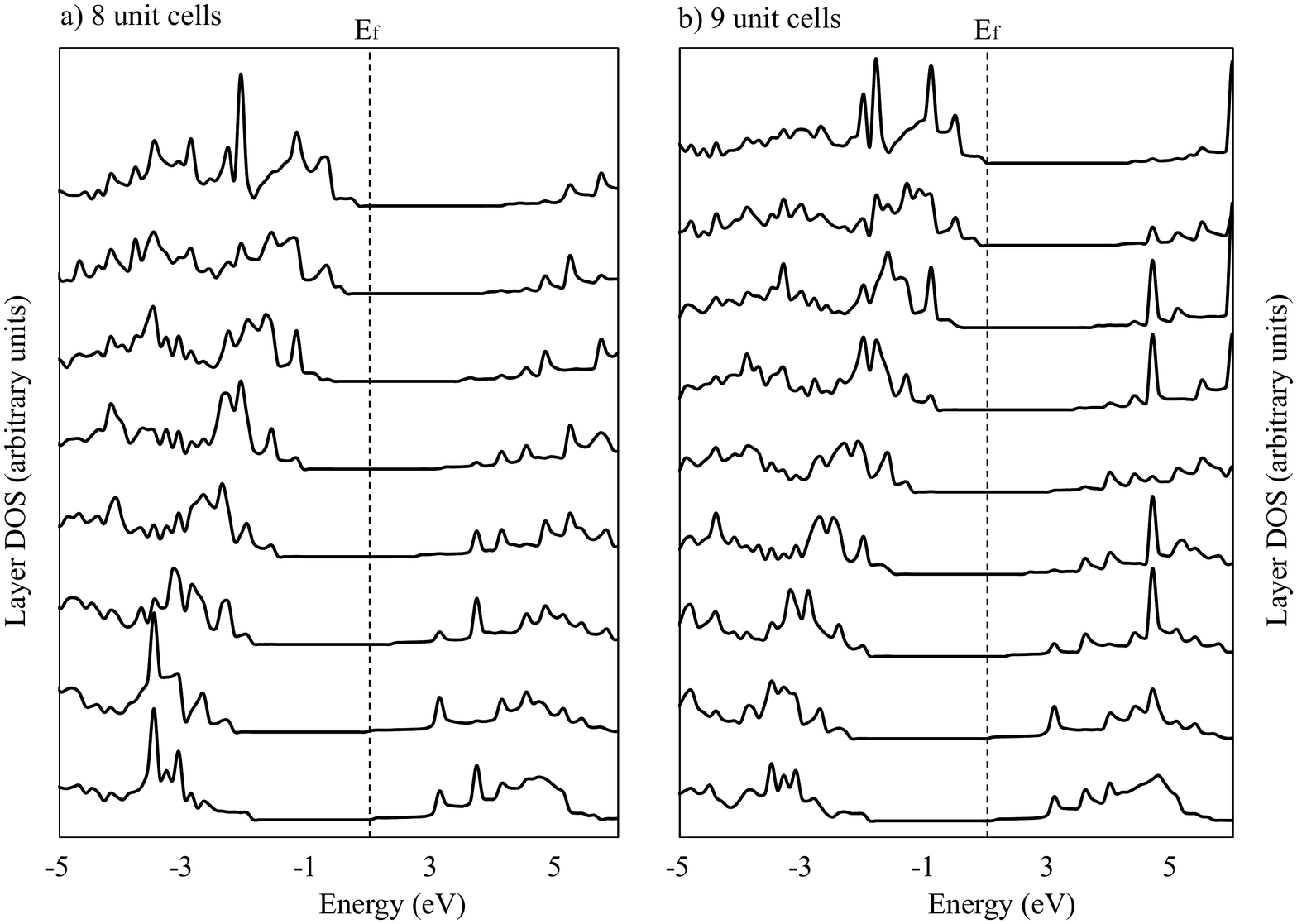}
    \caption{Layer-by-layer densities of states for slabs with a thickness of a) eight unit cells and b) nine unit cells, calculated using DFT. The slabs have KO top layers (with negative ionic layer charge) and TaO$_2$ bottom layers (with positive ionic layer charge). The Fermi level is set to 0 eV and indicated by the dotted vertical line. In the eight-layer slab the band bending is less than the band gap. In the nine-layer slab the surface densities of states cross the Fermi energy, indicating the onset of electrostatic screening through electron-hole excitation across the band gap. \label{fig:ldos_thickness}}
\end{figure}

In our DFT calculations of Fig.~\ref{fig:kto_thickness}, as well as the model of Fig.~\ref{minPFE_vs_d_figure}, we saw that P$_\xi$ has a constant saturation value below a thickness of 8 unit cells, and starts to reduce for larger thicknesses. We stated that the change in behavior is caused by the onset of screening through electron-hole excitation across the gap for larger thicknesses.
In Figs.~\ref{fig:ldos_thickness}a and b we show the layer-by-layer densities of states, calculated using DFT, for eight- and nine- -layer slabs respectively. For the eight-layer slab of Figs.~\ref{fig:ldos_thickness}a, we see that the band bending is less than the band gap, and neither the top of the valence band in the top layer, nor the bottom of the conduction band in the bottom layer intersect the Fermi energy, which is set to 0 eV.  
The layer-by-layer densities of states for the nine-layer slab of  Fig.~\ref{fig:ldos_thickness}b, reveal an increase in the band bending such that the bands in the surface layers cross the Fermi level. This results in compensating holes on the upper KO surface and compensating electrons on the lower TaO$_2$ surface.

\section{Detailed solution of Eqn.~\ref{Eqn:Etot}}
\label{Appendix:Etot}

We find the values of $P_{\xi}$ that minimize the total energy, $E_{\text{tot}}$, of Eqn.~\ref{Eqn:Etot} by taking the derivatives of $E_{\text{tot}}$ with respect to $\sigma$ and $P_{\xi}$, setting both to zero then solving the resulting equations simultaneously.

Differentiating with respect to $\sigma$ yields
\begin{equation}
    \frac{\partial E_{\text{tot}}}{\partial \sigma} = \frac{d}{\epsilon_0 \epsilon_r e} \left(-P_{\text{layer}} + P_{\xi} + \sigma \right) + \frac{E_g}{e} = 0
\end{equation}
which yields
\begin{equation}
    \sigma = P_{\text{layer}} - P_{\xi} - \frac{E_g \epsilon_0 \epsilon_r}{d} \quad .
\end{equation}
We recognize that, because of the finite size of the band gap, there is a critical thickness, $d_c$, below which compensating surface charges will not be created through this mechanism. The critical thickness is determined by
\begin{equation}
    P_{\text{layer}} - P_{\xi} \geq \frac{E_g \epsilon_0 \epsilon_r}{d_c}  \quad ,  \quad \text{or} \quad d_c \geq \frac{E_g \epsilon_0 \epsilon_r}{P_{\text{layer}} - P_{\xi}} \quad .
\end{equation}
Therefore we obtain
\begin{equation}
\begin{array}{llll}
\sigma & =  &  P_{\text{layer}} - P_{\xi} - \frac{E_g \epsilon_0 \epsilon_r}{d} \quad &  \text{for }  d \geq \frac{E_g \epsilon_0 \epsilon_r}{P_{\text{layer}} - P_{\xi}} \\
 & =  & 0 & \text{otherwise.} 
 \end{array} 
 \label{Eqn:sigma}
\end{equation}

Differentiating with respect to $P_{\xi}$ yields
\begin{equation}
    \frac{\partial E_{\text{tot}}}{\partial P_{\xi}} = \frac{d}{\epsilon_0 \epsilon_r e} \left(-P_{\text{layer}} + P_{\xi} + \sigma \right) + d (2a P_{\xi} + 4b P_{\xi}^3 ) = 0 \quad ,
\end{equation}
or
\begin{equation}
4b P_{\xi}^3 + \left(2a + \frac{1}{\epsilon_0 \epsilon_r e} \right) P_{\xi} + \frac{\sigma - P_{\text{layer}}}{\epsilon_0 \epsilon_r e} = 0 \quad .
\label{Eqn:Pxi}
\end{equation}

Substituting Eqn.~\ref{Eqn:sigma} for $\sigma$ in Eqn.~\ref{Eqn:Pxi} for $P_{\xi}$ we obtain
\begin{equation}
    \begin{array}{rcll}
  4b P_{\xi}^3 + 2a P_{\xi} - \frac{Eg}{de} & = & 0  \quad &  \text{for} \quad  d \geq \frac{E_g \epsilon_0 \epsilon_r}{P_{\text{layer}} - P_{\xi}} \\  
  4b P_{\xi}^3 + \left( 2a + \frac{1}{\epsilon_0 \epsilon_r e} \right) P_{\xi} - \frac{ P_{\text{layer}}}{\epsilon_0 \epsilon_r e} & = & 0  \quad &  \text{for} \quad  d < \frac{E_g \epsilon_0 \epsilon_r}{P_{\text{layer}} - P_{\xi}} 
    \end{array}
\end{equation}
We then find the roots of these equations using the Matlab given in Appendix~\ref{Appendix:Matlab}. 

\section{Matlab code for solving Eqn.~\ref{Eqn:Etot}}
\label{Appendix:Matlab}

{\tt
\begin{verbatim}
clear variables
close all

% abbreviations:
LW = 'linewidth'; FS = 'fontsize'; MS = 'markersize';
LOC = 'Location'; JL = 'JumpLine'; INT = 'Interpreter';
LX = 'latex';

%__________________________________________________________________________

volunitcell = (4.026e-10)^3;     % m^3
unitcellfactor = (4.026e-10)^2;  % m^2

%   Constants

PL = 0.5;             % The layer polarization in units of C/m^2
e = 1.6e-19;          % The electronic charge in C
eps0 = 9e-12;         % The permittivity of free space in C/V/m
apuc = -0.028;        % The a parameter in units of eV.m^4.C^{-2} per unit cell
a = apuc/volunitcell; % The a parameter in units of eV.m^4.C^-2 / m^3
bpuc = 1.123;         % The b parameter in units of eV.m^8.C^-4 per unit cell
b = bpuc/volunitcell; % The b parameter in units of  eV.m^8.C^-4 / m^3
Eg = 2.18;            % The band gap in eV

%   Parameters

epsr = 20;

dvec = 1.4e-9:0.5e-10:5e-8;
dvec = dvec';         % I prefer my list of d values to be a column vector
minPFEsigmavalid = NaN(length(dvec),1);  % Declare variables and fill them with NaN values 
minPFEsigmainvalid = NaN(length(dvec),1);
minPFEnosigmavalid = NaN(length(dvec),1);
minPFEnosigmainvalid = NaN(length(dvec),1);
smalldassumption = NaN(length(dvec),1);
largedassumption = NaN(length(dvec),1);

for i = 1:length(dvec)
    
    d = dvec(i);
    
    %   Small d limit calculation.  Sigma effect is assumed to be below the
    %   threshold and so sigma = 0 is used in the derivation of the
    %   derivative.  Note that this is actually independent of d.
    
    %fprintf('Small d minimisation:\n')
    smalldcoeffs = [4*b, 0, 2*a + 1/epsr/eps0/e, -PL/epsr/eps0/e];
    smalldderivroots = roots(smalldcoeffs);       % calculate the roots
    
    [~,ridx] = min(abs(imag(smalldderivroots)));  % work out which is the real root
    minPFEnosigma = smalldderivroots(ridx);       % this is the value of PFE that minimizes 
                                                  % the total energy when sigma = 0
    %fprintf('\n  PFE = %g is a minimiser\n',smalldPFEmin);
    
    %   Large d limit.  Sigma is asssumed to be large enough to have an
    %   effect so sigma is included in the derivation of the derivative
    %   and in the calculation of the minimising PFE.
    
    %fprintf('\n\nLarge d minimisation:\n')
    largedcoeffs = [4*b, 0, 2*a, -Eg/d/e];
    largedderivroots = roots(largedcoeffs);
    
    [~,ridx] = min(abs(imag(largedderivroots)));
    minPFEsigma = largedderivroots(ridx);
    %fprintf('\n  PFE = %g is a minimiser\n',largedPFEmin);
    
    %   We can check (after calculating the minimising PFE), the value of d
    %   that is needed for sigma to have an effect.  The assumption about the
    %   inclusion of sigma in the derivative calculation can then be tested.
    
    smalllimittest = eps0*epsr*Eg/(PL - minPFEnosigma);
    if (d < smalllimittest)
        %fprintf(' d = %g satisfies small d assumptions (d < %g)\n',d,smalllimittest);
        %smalldassumption(i) = 1;
        minPFEnosigmavalid(i) = minPFEnosigma;
    else
        minPFEnosigmainvalid(i) = minPFEnosigma;
    end
    
    largelimittest = eps0*epsr*Eg/(PL - minPFEsigma);
    if (d > largelimittest)
        %fprintf(' d = %g satisfies large d assumptions (d > %g)\n',d,largelimittest);
        %largedassumption(i) = 1;
        minPFEsigmavalid(i) = minPFEsigma;
    else
        minPFEsigmainvalid(i) = minPFEsigma;
    end
    
end

xscl = 4.026e-10;  % Convert the x-axis scale into number of unit cells
yscl = 100;        % Convert the y-axis scale into muC/cm^2

figure
plot(dvec/xscl,yscl*minPFEsigmavalid,'k-',LW,1.5)
hold on
%plot(dvec,minPFEsigmainvalid,'bo',LW,1.5)    (in case we want to look at the invalid values)
plot(dvec/xscl,yscl*minPFEnosigmavalid,'k-',LW,1.5)
%plot(dvec,minPFEnosigmainvalid,'ro',LW,1.5)  (in case we want to look at the invalid values)
hold off
ylim(yscl*[0.1,0.5])
xlim([0,max(dvec/xscl)])
%legend({'$\sigma$ effect (valid)','$\sigma$ effect (invalid)',...
%        'no $\sigma$ effect (valid)','no $\sigma$ effect (invalid)'},INT,LX)
%legend({'$\sigma$ effect (valid)','no $\sigma$ effect (valid)'},INT,LX)
xlabel('Number of unit cells',INT,LX)
ylabel('$P_{\xi}~$   ($\mu$C/cm$^2$)',INT,LX)

set(gca,'TickLabelInterpreter','latex')  
set(gca,'fontsize',14,'FontWeight','bold')

print -deps 'minPFE_vs_d_figure.eps'
\end{verbatim}
}

\newpage
\bibliographystyle{unsrt}
\bibliography{references,Nicola}
\end{document}